# Depth and Breadth of Research Area Coverage and Its Impact on Publication Citation: An Analysis of Bibliometric Papers

Zhuoran Lin[a*], Yun Wang[b, c*], Hongjun Li[a,1]

[a] *Library, China Agricultural University, Qinghua Donglu No. 17, Haidian District, Beijing 100193, China*

[b] *College of Humanities & Development Studies, China Agricultural University, Qinghua Donglu No. 17, Haidian District, Beijing 100193, China*

[c] *Human Resource Development Center of MARA, China Association of Agricultural Science Societies, Maizidian Street No. 22, Chaoyang District, Beijing 100125, China*

**Abstract**

Many other factors affecting citation of publications, except for research area coverage, have been studied. This study aims to investigate impact of research area coverage. Bibliometric papers and their related papers (referred papers, citing papers and first author's papers) were screen and matched by Python program. Papers' research areas were classified according to Web of Science. Bibliometric parameters of the most cited 5% and the least cited 5% papers were compared. Firstly, coverage of related papers' research areas impacts the citation of their original papers. The impact of references and citing papers are positive and negative, separately, while the first author's papers have no influence. Secondly, high-influence papers tend to cite references from a wider area and are cited by followers from a wider area. Additionally, the pattern of knowledge flow differs significantly between high- and low-influence papers. Low- influence papers narrow knowledge flow, whereas high-influence papers broaden it. This study has shown that both depth and breadth of research area coverage can influence citations. It is recommended that authors should extensively cite high-influence publications, both within and beyond their own area.

**Introduction**

Publishing is a part of researchers' life, because publications represent good results of their research, increase exposure times to colleagues, and earn them more reputation on scientific community. Furthermore, it plays crucial role in many aspects for researchers, such as promotion, reputation, salary and job mobility (Miller, 2011). It can't be denied that publication is an arena for researchers. Just as comparing the skill of fishermen by the weight of their catch rather than the quantity, evaluating researchers' performance should balance the quality and quantity of their publications.

For papers, cited times is the original and most important indicator to evaluate their impact.

[1]  E-mail addresses: 1124230606@qq.com (Zhuoran Lin), nxhkczg@126.com (Yun Wang), lihj@cau.edu.cn (Hongjun Li, corresponding author, ORCID 0000-0002-9048-1420)
*These two authors contribute equally.

Although citation indexing was originally proposed for information retrieval services, it is widely used to evaluate publication impact shortly after its birth. For instance, high-quality articles were found to be cited more often (Patterson & Harris, 2009). Grayson et al. (2021) used citation analysis to evaluate most impact papers in the field of pediatric neurosurgery. Citations are criticized for being one-dimensional and merely measuring the impact of science on the scientific community, but it is still the core of all indicators, because other indicators are derived from citations. For example, h-index was defined as the number of papers with citation number ≥ h by Hirsch (2005) to measure the impact of a researcher's publications, and was modified by others (Alonso et al., 2009; Egghe, 2010; Poirrier et al., 2021) to address some limitations. However, the assessment of publication impact based solely on citations, whether from literature or social media, is not entirely reliable. Nevertheless, it cannot be disregarded, especially when papers are not subject to rigorous review. As a result, bibliometric research on publication cited times has been a popular topic of discussion for decades. Such research seeks to improve the accuracy of measuring publication impact, and to develop more effective strategies for evaluating the quality and significance of scholarly work.

**Literature review**

The number of citations is a seemingly straightforward metric, but it is actually influenced by a multitude of interrelated factors, such as journal influence, title length, text length, author numbers, references numbers, collaboration type, language, etc (Xie et al., 2019). These factors can be broadly categorized into three groups: paper-related factors, author-related factors, and journal-related (Tahamtan et al., 2016). Of these, the quality of the paper undoubtedly plays the most significant role in determining its citation count (Jabbour et al., 2013). This has been supported by research demonstrating a statistically significant correlation between expert evaluations of paper quality and citation counts (Buela-Casal and Zych, 2010). Additionally, factors such as the novelty of the method, design, and results, as well as the current popularity of the topic, can also impact citation counts. For example, papers that propose innovative connections between clusters of co-cited references are more likely to garner citations in the future (Chen, 2012).

Among several journal-related factors, the journal impact factor is considered to have the most significant influence. Numerous studies have confirmed a positive correlation between JIF and citations, including research by Vanclay (2013), Shuaib et al. (2015), Sun et al. (2020), and Traag (2021). However, some studies do not support this conclusion (Roldan-Valadez and Rios 2015; Kulczycki et al. 2021). Roldan-Valadez and Rios (2015) and Kulczycki et al. (2021) don't agree that either the impact factor of citing journals or the size of cited journals is a good predictor of the number of citations, because there are questionable journals on lists of Web of Science or Scopus.

Several author-related factors have been extensively studied in the literature, including the number of authors, the reputation of the authors, and collaboration between authors. Studies have suggested that papers with more authors and a stronger reputation are more likely to be cited (Bjarnason et al., 2002; Fox et al., 2016; Abt, 2017). Sooryamoorthy's (2017) analysis of social science publications in South Africa revealed that research with collaborative efforts tended to receive more citations than research without such efforts. Other author-related factors also have influence on citations, such as authors' productivity (Bornmann and Daniel, 2007), authors' country (Peng and Zhu, 2012), and reputation of authors' organiztions (Amara et al., 2015).

Sjögårde and Didegah (2022) introduced a novel factor, topic growth, for citation evaluation. Their study reveals that publications in rapidly developing topics receive more citations compared to those in declining or slow-growing areas. This finding is in line with previous research examining the higher citation rates of emerging research topics (Kwon, et al. 2019; Thelwall and Sud, 2021).

Highly cited papers are often the most sought-after among scholarly works, as their elevated citations are indicative of significant impact and the potential for authors to be regarded as top scientists. A compelling statistic supports this notion, as 59 Clarivate Analytics Citation Laureates have gone on to win Nobel Prizes as of 2021 (Clarivate, 2021). Consequently, recent years have seen a proliferation of studies investigating the characteristics of highly cited papers and the reasons why they garner more citations (Perez-Cabezas et al., 2018; Hou et al., 2020; Di Zeo-Sánchez et al., 2021). In general, highly cited papers are distinguished by their novelty, significance, diversity, and at times, disruptive results.

A paper, like a journal, has a coverage of field, subject, research area or topic (hereinafter referred to as "research area"). If references of one paper cover many research areas, it is considered to be a diverse source of knowledge and is often perceived as having a broad research area. Conversely, if most of a paper's references come from one or two research areas, it is seen as concentrating knowledge on specialized research areas, resulting in a narrow but deep coverage of the research area. This coverage can also be determined by the research area of the citing papers or the author's own papers.

Despite extensive research on various factors, research area coverage has received relatively little attention. However, some studies have shown that the research area coverage of a paper or journal can influence citations. For example, papers published in general journals like *Nature* and *Science* are expected to receive more citations than those published in specialized journals, as these journals have a broader readership (Vanclay 2013). Similarly, papers covering narrow topics or in small fields may receive fewer citations than those covering more extensive and general fields.

To predict future citations, Chakraborty et al. (2013 and 2014) introduced the diversity of references and authors' papers as important indicators, because interdisciplinary research is believed to result in better outcomes and to have more followers.

As previously mentioned, there have been several studies examining the impact of references, citing papers, and authors' papers on citations. However, most of these studies focus on visible factors, such as the number of papers, publication year, and the cited times of references and citing papers. For example, Mammola et al. (2021) found that papers with a higher number of references, high impact and recent references, tend to receive more citations. Nevertheless, to the best of our knowledge, no study has investigated the effects of reference, citing publications, and author's publications on bibliometric papers' citations. Moreover, it remains unclear whether citing more bibliometric documents or being cited by more bibliometric documents positively influences citations.

This paper aims to analyze the relationship between research area coverage and cited times, using bibliometric papers as samples. Specifically, we aimed to address the following questions:

Q1. Is the research area coverage of references, citing publications, and first-author's publications similar?

Q2. Does the research area coverage of references, citing publications, and first-author's publications affect cited times? Which one plays a stronger influence?

Q3. Which has a greater impact on cited times – the width or depth of area coverage?

Q4. How does the area coverage of bibliometric publications change from cited papers to citing papers?

**Methods**

Papers in core database of Web of Science are categorized into three levels of research areas: macro, meso, and micro topics. These levels consist of 10 macro topics, 326 meso topics, and 2441 micro topics. To eliminate any ambiguity, we refer to them as macro research area, meso research area, and micro research area. For example, bibliometric is a meso research area, which is classified into macro research area (social sciences), and there are three micro research areas in bibliometric.

In this study, 16975 bibliometric papers (only including research articles and review articles) published between 2012 and 2021 in Web of Science core are retrieved in January 2023, and then are cleaned, analyzed following the procedure (Figure 1). The measurement of research area width and depth is based on the percentage of papers, making it vulnerable to unreasonable deviations from papers with few references or citations. For instance, while one paper with only two references (one of which is from area bibliometric) and another paper with ten references (five of which are from area bibliometric) both have a 50% area bibliometric reference percentage, they significantly differ in measuring area coverage. Consequently, to ensure greater accuracy and reliability, future steps will exclude papers with less than ten citations or references.

Because the very new papers receive less citations than papers published several years ago, all papers are ranked by citations from high to low in their published year, and samples are selected. It is obviously that the top 5% papers are the most cited papers and the last 5% papers are those with cited times of 10 or a little higher.

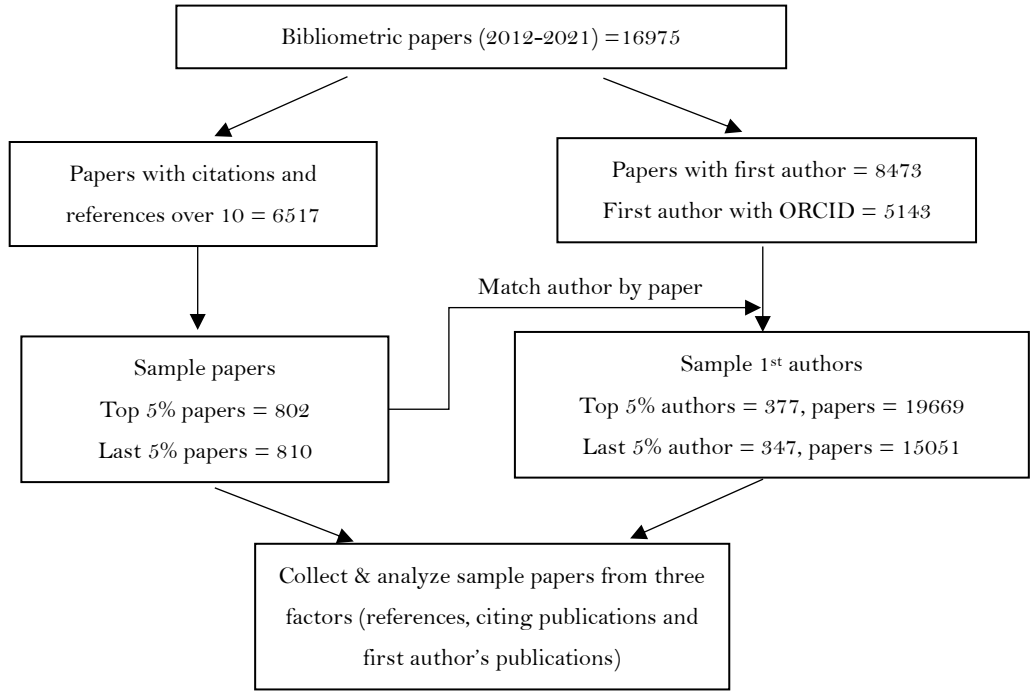

Figure 1: Procedure for collecting and sampling papers

To ascertain the influence of the author on citations, we solely analyze papers of the first authors, as they make the most significant contribution to the publication and possess more sway over the other authors. There are 5143 distinct authors in 8473 papers with initial author ORCID, and these authors are sorted into groups by matching them with sample papers. In the case that an author's publications belong to multiple groups, the higher-ranking group takes precedence. For instance, if an author's papers are both in the top 5% and the last 5% groups, he or she is categorized as the top 5% author.

The Shannon Index, which has been in use since its introduction in 1948 (Shannon 1948a & 1948b), is widely employed to gauge community diversity. To get diversity of publications, the entropy of references, citing publications and author's publications is employed as the following formula (Chakraborty et al.2014)

$$\text{Diversity} = \sum_{i=1}^{n} \left(\frac{X_i}{X}\right) * \ln\left(\frac{X_i}{X}\right)$$

where $X_i$ denotes the number of papers belonging to area $i$, $X$ denotes the total number of papers, $X_i/X$ means the percentage of papers in area $i$, and n is the number of involved areas. Diversity is equal or larger than 0, and a higher diversity represents more diversity of groups.

To evaluate whether references, citing publications or author's publications from a specific area impacts citations, we introduce the degree of majority which is dependent on the percentage and ranking of papers from one area. The formula for the degree of majority is as

$$\text{Degree of majority} = \left(\frac{Xi}{X}\right) * \frac{1}{Ri}$$

where $Xi$ represents the number of papers belonging to area i, X represents the total number of papers, and $Ri$ is the ranking of area i. This formula comprises two parts, percentage and ranking, and accurately reflects the trend of majority, where an area with a greater percentage and higher ranking will have a higher degree of majority. The two extremes of degree of majority are 0 (one area without any papers) and 1 (one area covers 100% papers). When two areas have the same value of percentage or ranking, degree of majority is determined by the other part of the formula. For example, if area A and area B (in different papers) both cover 30% of the papers and rank No. 1 and No.2. And their degrees of majority are 0.3 and 0.15 using this formula, indicating that area A is more important than area B. Similarly, area C and area D (in different papers) both rank second and cover 30% and 29% of the total papers. Their degrees of majority are calculated as 0.15 and 0.145, respectively, thereby indicating that area C has slightly more priority than area D.

All the data were processed using Microsoft Excel. The statistical significance was determined by Duncan multiple range test was conducted by SPSS v21.

**Results**

*General overview*

In 2012-2021, 16975 bibliometric papers are published in 3220 journals, and the top 5% and the last 5% papers are published in 287 journals and 388 journals, which covers 8.9% and 12.0% journals, respectively (table 1). *Scientometrics* publishes 15% papers, which is far more than the following 2 journals, *Journal of Informetrics* (4.3%) and *Plos One* (2.9%). In group the top 5%, *Plos One* jumps to 2nd largest journal, while in group the last 5%, *Journal of the Association for Information Science and Technology* replaces *Plos One* to be the 3rd largest journal. The result shows that *Plos One* is the most source of bibliometric papers besides traditional library and information science journals, and the bibliometric papers in *Plos One* might get more citations.

The average cited times of all papers, the top 5% and the last 5% papers is 15.2, 108.3 and 10.8, respectively, which is higher than the median of cited times. Similarly, the average numbers of references are also higher than medians. Therefore, the distribution of cited times and reference numbers shows a left skewness, which means most papers get less citations and have less references than the average.

| Paper type | Number of Paper | Cited Times | | Number of Reference | | Number of Journal | Top 3 journals of most papers |
|---|---|---|---|---|---|---|---|
| | | average | median | average | median | | |
| Overall | 16975 | 15.2 | 7 | 37.6 | 32 | 3220 | Scientometrics (15.0%) Journal of Informetrics (4.3%) Plos One (2.9%) |

| | | | | | | |
|---|---|---|---|---|---|---|
| Top 5% | 802 | 108.3 | 79 | 56.8 | 47 | 287 | Scientometrics (14.0%)<br>Plos One (7.0%)<br>Journal of Informetrics (6.8%) |
| Last 5% | 810 | 10.8 | 10 | 44.6 | 37 | 388 | Scientometrics (16.2%)<br>Journal of Informetrics (4.8%)<br>Journal of the Association for Information Science and Technology (3.1%) |

Table 1: Statistical overview of papers

Note: Percentage following journal name is paper proportion of journal in dataset.

When examining papers sorted by their bibliometric percentage, it becomes clear that there is a significant difference in the distribution of reference, citing paper and author paper. Specifically, references tend to exhibit an inverted U-shape, indicating that most papers possess a medium citation strength (ranging from 20% to 60%) in terms of their bibliometric publications (Figure 2). On the other hand, the distribution of author papers presents a U-shape. In other words, bibliometric publications constitute only a small part of most authors' papers, suggesting that authors tend not to limit themselves to one particular area of research. For example, 31.6% of authors publish less than 10% bibliometric papers in relation to their overall output, while 15.3% and 9.7% of authors produce bibliometric papers that make up 10-20% and 20-30% of their total work, respectively. Meanwhile, there are 8.4% of authors whose bibliometric paper percentage exceeds 90%, indicative of their concentrated focus on bibliometric research. The coefficient of variation for citing papers is 19%, which is much lower than both references (coefficient of variation= 38%) and authors (coefficient of variation= 78%). Furthermore, when the highest and lowest groups are excluded (each consisting of 13.3% and 6.0% of the total sample, respectively), the coefficient of variation drops to 11%. These findings suggest that the distribution of citing papers is more stable than that of references and authors.

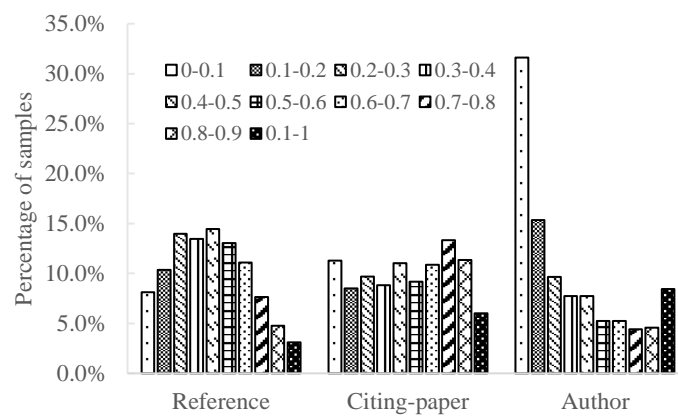

Figure 2: Histogram of samples of reference, citing-paper and author-paper by bibliometric percentage

Note: Number with legend represents the range of bibliometric percentage.

Area-related factors include width and diversity of research area. When considering the depth of the area, it is important to take into account the percentage of the specific area, such as area bibliometrics, in this study. Through correlation analysis, it has been discovered that three factors from citing papers, namely the number of meso areas, number of bibliometric papers, and diversity of paper, as well as three factors from references, including the number of bibliometric papers, percentage of bibliometric paper, and diversity of paper, have a significant relationship with cited times (Table 2). Two factors from citing papers, the number of meso areas and the number of bibliometric papers, are strongly and positively correlated with cited times, with correlation coefficients of 0.827 and 0.739, respectively. On the other hand, the other three factors, namely the diversity of citing paper, number, and percentage of bibliometric paper, have a medium to weak positive correlation with cited times.

| Paper dataset | Factors | Pearson coefficient | Significant |
|---|---|---|---|
| Citing publications | Number of meso area | 0.827 | 0 |
| | Number of bibliometric paper | 0.739 | 0 |
| | Percentage of bibliometric paper | -0.04 | 0.108 |
| | Diversity of paper | 0.386 | 0 |
| References | Number of meso area | -0.004 | 0.882 |
| | Number of bibliometric paper | 0.254 | 0 |
| | Percentage of bibliometric paper | 0.270 | 0 |
| | Diversity of paper | -0.077 | 0.002 |

Table 2: Correlation analysis between cited times and area-related factors

*Impact of References*

The average reference count for the top 5% is 56.8, significantly higher than the last 5% (table 3). Roughly one-tenth of the top 5% cites over 100 references, with an average citation count of over 140. This is ten times higher than the overall average citation count and 40% higher than the top 5% on average. The top 5% also has twice the percentage of highly cited papers in their references, indicating a preference for citing other top papers. The top 5% and the last 5% have a similar range of meso area references (1-27), but the top 5% covers more meso areas (5.8) than the last 5% (5.1), signifying a greater breadth of knowledge. Interestingly, the meso area diversity of the top 5% and the last 5% is very close, indicating little effect of reference diversity on citations.

Most papers tend to cite more articles within their own area, given the shared background, methods, and results. In sample of 1612 papers, 83.7% have bibliometric articles as their primary source of reference, with an additional 10% choosing area bibliometric articles as their secondary source. The bibliometric reference proportion of the top 5% and the last 5% is 48.43% and 39.63%, respectively, with a significant difference between them. The top 5% citing area bibliometric articles as their primary source of reference have an average citation count of 116.1, higher than the overall the top 5% average of 108.3 and the rest of the top 5% average of 77.6. There are 58 papers that do not cite any references from area bibliometric, 25 of which are from the top 5%, but their citation count (37.8) is much lower than the rest of the top 5% with references from area bibliometric (111.5). These results suggest that bibliometric papers can benefit from prioritizing bibliometric references to improve their citations.

| Paper type | Number of Reference | Percentage of Highly cited papers | Number of Meso area | Diversity_Meso area | Percentage of Bibliometric papers |
|---|---|---|---|---|---|
| Top 5% | 56.8 | 7.13% | 5.8 | 0.927 | 48.43% |
| Last 5% | 44.6 | 3.60% | 5.1 | 0.916 | 39.63% |
| Sig. | <0.01 | <0.01 | <0.01 | - | <0.01 |

**Table 3: Comparison of references of sample papers**

It cannot be overlooked that some peculiar papers with minimal references receive an unexpectedly large number of citations. For instance, six papers have been cited over 300 times (422, 551, 551, 1258, 390, and 308), and yet they have less than 30 references each (20, 23, 26, 26, 30, and 30, respectively). All of them are research articles and published in top journals of interdisciplinary or information science and library science: 2 in *Scientometrics*, 2 in *Science*, 1 in *Proceedings of the National Academy of Sciences*, 1 in *Journal of the Association for Information Science and Technology*. These papers predominantly focus on bibliometric as their primary area of research, with the proportion of bibliometric references ranging from 61% to 100%. One PNAS paper titled "*Rescuing US biomedical research from its systemic flaws*" cites 20 publications, with four of them from the bibliometric field, constituting 20% of all references. However, it is worth noting that 11 references in this paper are books or dissertations that cannot be classified under any discipline area, so the bibliometric percentage increases to 44.4% when non-papers are excluded.

Although 25 papers without any bibliometric reference are listed in the top 5%, their average citation count is a mere 37.8, which is approximately 1/3 of the top 5%. In contrast, out of 16 papers with all references sourced from the bibliometric field, 14 are in the top 5%, with an average citation count of 249.6, which is 2.5 times that of the top 5%. These findings further support the notion more bibliometric references improves citations.

*Impact of Citing Publications*

The top 5% and the last 5% is cited by 87.7 and 8.7 publications on average, which is 10% less than cited times (108 and 10.8) due to multiple citations from the same source (table 4). Within the citing publications, the top 5% has a significantly higher percentage of highly cited papers than the last 5%, indicating a mutual benefit of cited times between the citing and the cited publications. Notably, the top 5% covers a wider range of meso areas (17.5 areas with a diversity 1.637) than the last 5% (3.5 areas with a diversity of 0.910), emphasizing that the number and diversity of citing publications can enhance citations. The proportion of bibliometric papers in the top 5% is 47.11%, which is lower than the last 5% by 6.6 percentage points, and the difference is significant.

| Paper type | Number of Citing | Percentage of Highly cited | Number of Meso | Div_Meso area | Percentage of Bibliometric |
|---|---|---|---|---|---|

|        | publication | papers | area  |       | papers |
|--------|-------------|--------|-------|-------|--------|
| Top 5% | 87.7        | 1.87%  | 17.54 | 1.637 | 47.11% |
| Last 5%| 8.7         | 0.85%  | 3.47  | 0.910 | 53.72% |
| Sig.   | <0.01       | <0.01  | <0.01 | <0.01 | <0.01  |

**Table 4: Comparison of citing publications from sample papers**

Comparing the citing publications and cited publications, it is evident that the top 5% cites more references, has more citations, wider area coverage, and more paper diversity than the last 5%, for both citing publications and cited publications. However, the percentage of bibliometric publications is substantially different between the top 5% and the last 5%, being higher for the top 5% in citing publications and lower in cited publications (Figure 3). The top 5% receives 48% of knowledge from bibliometric areas and contributes 47% to it, indicating a stable knowledge flow. In contrast, the last 5% obtains over 60% of its knowledge from outside bibliometric areas but shares only 47% of it. This suggests a knowledge flow pattern of "wide input and narrow output" for the last 5%.

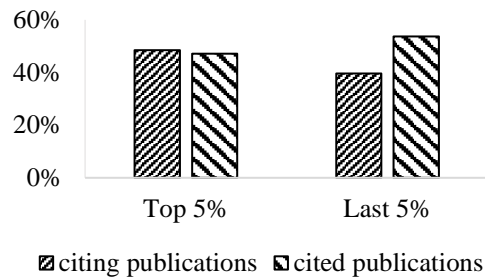

Figure 3: Proportion changes of bibliometric publication from citing- to cited-paper

In this section, 5 highest percentage of bibliometric citation papers and 5 lowest percentage of bibliometric citation papers from the top 5% with cited times over 200 are selected to analyze the effect of extreme proportion of bibliometric on cited times. The percentage of bibliometric citing publications of the 5 highest percentage of bibliometric citation papers ranges from 86.4% to 94.7%, while that of the lowest percentage of bibliometric citation papers ranges from 4.71% to 11.23%. Four out of the five highest percentage of bibliometric citation papers are published in traditional library and information science journals (2 in *Scientometrics*, 2 in *Journal of the Association for Information Science and Technology*), while one is published in *BMC Medicine*, which is medical journal. The lowest percentage of bibliometric citation papers, on the other hand, have only one paper published in a traditional library and information science journal (*Journal of Informetrics*), with the other four published in multidisciplinary and specific journals (2 in *Plos One*, 1 in *Journal of Business Research*, and 1 in *Expert Opinion on Biological Therapy*). Despite their varying publication sources, all these papers focus on traditional topics of bibliometric, scientometrics, knowledge mapping, and peer review, making bibliometric the primary source of citations, except for the one paper in *Journal of Business Research* with a focus on management. The paper titled "*How to conduct a bibliometric analysis: An overview and guidelines*" has the highest number of followers (89 papers, 28 of which are from its own journal), and bibliometric ranks second in terms of the number of papers cited (29 papers), but still far higher than third area (17 papers). In non-bibliometric journals, self-citations often prevent bibliometric citations from becoming the primary source, but when it comes to bibliometric-specific topics, bibliometric citations remain the most prominent.

*Impact of First author's papers*

While the productivity of the top 5% first author is higher at 52.2 papers than the last 5% first author's 43.4 papers, the difference is not statistically significant (Table 5). Within the top 5% group, the productivity of the top 10 authors ranges from 252 to 994 papers with an average of 531.8, while the range for the same metric is between 274 and 537 papers with an average of 426.5 in the last 5% group. The inverse relationship of productivity between all authors and top 10 authors suggests that some extreme yield authors may influence the apparent productivity of the last 5%. Removing 10 highest yield authors shows a significant difference in average productivity between the top 5% (42.0 papers) and the last 5% (28.9 papers) authors, with a p-value <0.01.

In the last 5% group, only 1 of the 10 highest yield authors focuses on bibliometric, accounting for 9.7% of their papers. The percentage of bibliometric publications for the other nine authors is no more than 7.7%, except for one with a percentage of 14.6% (although the No. 1 area percentage is 51.8%). In contrast, in the top 5% group, 4 of the 10 highest yield authors publish 63.6-94.6% of their papers in the area of bibliometric, and 2 authors focus on bibliometric as their No. 2 area, each of which accounts for 20% and 3.4% of their bibliometric coverage, respectively. This clearly shows that authors in the top 5% group prioritize bibliometric research, while those in the last 5% group do so much less frequently. This finding is supported by the bibliometric percentages of the entire the top 5% and the last 5% groups, as the top 5% authors publish 36.40% of their papers in the area of bibliometric, while the last 5% authors publish only 33.25% of their papers in this area.

| Paper type | Number of Paper | H-index | Percentage of Highly cited papers | Number of Meso area | Div_Meso area | Percentage of Bibliometric papers |
|---|---|---|---|---|---|---|
| Top5% | 52.2 | 16.4 | 2.92% | 9.0 | 1.273 | 36.40% |
| Last 5% | 43.4 | 10.7 | 0.91% | 8.7 | 1.324 | 33.25% |
| Sig. | - | <0.01 | <0.01 | - | - | - |

**Table 5:** Comparison of first author's papers of sample papers

As expected, authors in the top 5% group have a significantly higher H-index and higher percentage of highly cited papers than authors in the last 5% group (Table 5), as both indicators depend on both the quantity and quality of publications. Unlike the significant difference of meso area coverage and diversity between the top 5% and the last 5% of citing and cited publications, the difference in the authors' publications is relatively small. Compared to 17.5 meso areas of citing publications in the top 5% group, 9 research areas of interest for the authors are rather low but still higher than 5.8 meso areas of references in the top 5% group. Paper diversity of authors is around 1.3, which is between that of citing papers (1.6) and references (0.9). The percentage of bibliometric papers of authors is 34.9%, which is lower than that of citing publications (50.4%) and references (44.0%). This can be explained by the fact that authors have research interests beyond bibliometric. Although bibliometric is the favorite research area for 40.9% of authors, it is ranked lower than No. 4 area by 27.2% of authors. Area management and area knowledge engineering & representation are the second and third most popular areas for authors due to the close relation to bibliometric.

*Combined impact of references, citing papers and first author's papers*

The distribution of references, citing papers, and author's papers in the top 5% varies significantly (Figure 4). For references, the percentage of the top 5% increases steadily as the proportion of bibliometric goes up, with the top 5% papers surpassing the last 5% papers by 5-10% in medium proportion of bibliometric (0.4-0.8), and rapidly increasing to 90% in high proportion of bibliometric (0.8-1). However, for citing papers, the percentage of the top 5% decreases slowly from 60% in low proportion to 27% in high proportion of bibliometric. For author's papers, the percentage of the top 5% author's papers is around 50% with slight fluctuations when proportion of bibliometric is between 0 and 0.7, with an increase to 82% at bibliometric proportion of 0.8-0.9, followed by a sharp decline to 41% when bibliometric proportion reaches 0.9-1. These lines of distribution clearly indicate that citing more references from the area of bibliometric is helpful to citation of a bibliometric paper, and that more citing papers from its own area may bring some negative effects on cited times. Author's coverage of interesting has little influence on citation, except for an author with an extreme concentration on bibliometric.

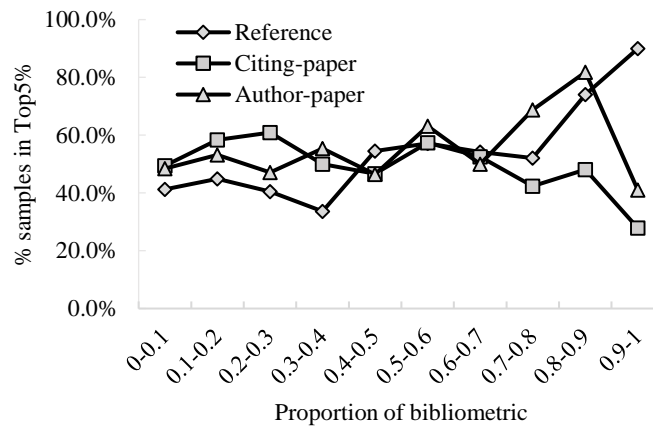

Figure 4: Distribution of samples in the top 5% by proportion of bibliometric paper

In the sections above, we discuss the impact of area coverage of citing and cited publications on citation. As part of a paper, references, citing papers and authors might interact on citation. The average degree of majority of bibliometric references is 0.429, with degree of majority of 5 small groups no more than 0.422, while degree of majority of 5 large groups is 0.470 (Figure 5). However, the degree of majority of bibliometric citing publications has a different effect on citations. The average degree of majority of 5 small groups is 0.518, which is larger than that of 5 large groups. Additionally, statistical analysis demonstrates a significant difference of reference degree of majority between the top 5% (0.471) and the last 5% (0.387) and the same for citing publications (0.453 for the top 5% and 0.516 for the last 5%). By the way, the degree of majority of the top 5% author and the last 5% author is 0.32 and 0.281, respectively, but there is no significant difference. In conclusion, the coverage of areas indicates that references have a more crucial role in cited times than citing publications, while the research area of authors has little influence.

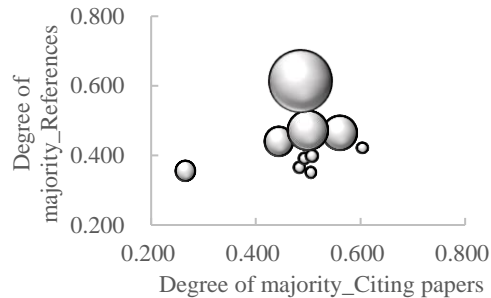

Figure 5: Relationship between cited times and degree of majority of citing papers and references

Note: a) 1612 papers are divided into 10 groups by cited times, and b) bubble size represents average cited times of each group.

Bibliometric papers tend to cite approximately 44% of their own area's publications and have about 51% of their followers from the same field. According to the theory of knowledge flow, the breadth of knowledge narrows with a decrease of 7% from references to citing publications as the proportion of non-bibliometric sources becomes smaller. Specifically, papers with lower citations receive a greater net flow of knowledge from bibliometric sources compared to those with higher citations (Figure 6). For instance, eight groups (group 1-5 and group 7-9) show a positive net flow of bibliometric knowledge, which limits their influence on the area of bibliometric. This is especially notable for group 1 and 2, where the percentage of non-bibliometric sources decreases by almost 20%, while group 3-5 and group 8 have a decrease of about 10%. In contrast, group 6 with 30 citations and group 10 with 259 citations garner more attention beyond the realm of bibliometric than they receive from non-bibliometric sources. In group 10, while the percentage of references from the area of bibliometric is 62.2%, only half of its followers are from within this field. This indicates that influence of highly cited papers is not limited to their own area but also extends into non-bibliometric domains.

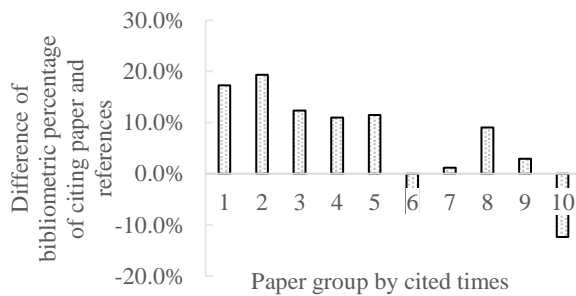

Figure 6: Knowledge flow of bibliometric from reference to citing publication

Note: a) 1612 papers are divided into 10 groups by cited times (1 is the lowest and 10 is the highest), and b) bibliometric percentage in vertical axis is value of citing paper minus value of reference.

At the meso level, the area of bibliometric serves as the primary source that papers follow and the main destination for papers to exert influence. Social sciences, which is the largest macro area among 10 areas and the parent level of bibliometric, accounts for 79.5%, 73.1%, and 60.9% of the cited, citing, and first-author publications, respectively. Clinical & life sciences and electrical engineering, electronics & computer science are the second and third largest macro areas, respectively. Clinical & life sciences provides 10.3% of cited publications, 13.2% of citing publications, and 19.7% of first-author publications, possibly because it is the most prominent source of papers in the Web of Science. Meanwhile, electrical engineering, electronics & computer science comprises approximately half of clinical & life sciences, as bibliometric has a close relationship with this field. For instance, librarians on information science often learn from computer science to enhance library performance. The remaining seven macro areas each cover less than 10% of the publications for bibliometric documents. In summary, social sciences play a crucial role as a major source of knowledge and responders to bibliometric documents, while the area coverage from input to output continues to expand.

**Conclusions**

As a less prominent research direction, bibliometric papers are primarily published in library and information science journals, where *Scientometrics* is the largest source journal and accounting for one-sixth of total publications in the bibliometric area. Besides library and information science area, open access and medical journals serve as another key sources. As far as knowledge inheritance and flow are concerned, references and citations are primarily reside within the same research area of publications, which is consistent with earlier studies (Hammarfelt, 2011; McLevey et al., 2018; Zhang et al., 2021). In addition to common characteristics of bibliometric publications, this study has uncovered new findings.

Firstly, it is important to note that paper distributions vary greatly depending on bibliometric proportions. In terms of citing publications, papers are distributed fairly evenly with minor fluctuations, while for references and author publications, there is a clear contrast. An even distribution suggests no preference for bibliometric areas, as all bibliometric papers receive similar recognition within and beyond their own fields. A U-shaped distribution, on the other hand, signals that two extremes (favoritism or disinterest in bibliometric areas) have become dominant. This, in turn, leads to the identification of two groups of authors: bibliometric-focused authors and extensive-interest authors. Interestingly, the bibliometric-focused group is much smaller than the extensive-interest group, highlighting the trend of authors with broad research areas in this era of knowledge explosion and cross-disciplinary collaboration. In terms of references, an inverted U-shaped distribution suggests that most papers prefer to cite publications from their own area, given that knowledge within the same field is more easily shared and understood.

Secondly, research area coverage significantly affects publication citations. In terms of references, the coverage of bibliometric has a positive impact on cited times, while citing papers have a negative impact. However, the research coverage of the author does not have a significant influence on cited times. A paper with more references from the bibliometric area indicates that it has obtained more knowledge from the field and positions itself as a member of the bibliometric community. With more bibliometric features in a publication, a paper becomes easier to find among millions of publications by readers interested in the field and preparing to write bibliometric articles. The distribution of followers demonstrates the range of impact. If a paper has only or mostly followers from the bibliometric area, it may have less impact on other areas. In the fast-developing field of interdisciplinary research, a paper may gain more attention through widespread channels, leading to an increase in the number of citations received. The author's research interest is not a critical factor in determining citations, as anyone with a good idea, even without prior experience in bibliometric research, can write an excellent bibliometric article. With the aid of the internet, authors without experience in bibliometric research can easily acquire information and quickly master the research paradigm of the bibliometric area before preparing for their paper.

Thirdly, the breadth of research area coverage has a significant impact on citations for both cited and citing papers. The results indicate that top papers not only attract more readers, but also follow wide publications for a variety of purposes, including background information, citing methodology, arguing opinions, and demonstrating results (Rong and Safer, 2008; Stefan et al., 2015; Dorta-Gonzalez, 2019). Papers with high influence and broader coverage tend to have wider dissemination, resulting in a higher likelihood of citation. Broader coverage typically implies a more diversified range of publications and stands in contrast to depth. Although the diversity and breadth of cited and citing publications of high influence papers is better less impact papers, the depth of bibliometric references does not consistently align with diversity or breadth. Furthermore, the diversity, depth, and breadth of an author's publications have little impact on citations.

Fourthly, the pattern of knowledge flow differs between high and low influence papers. Low influence papers tend to focus on traditional topics within their area of bibliometric, and their impact is largely limited to their specific area, resulting in a narrow function of knowledge dissemination. On the other hand, high influence papers not only promote knowledge dissemination within their area of bibliometric, but also across the entire scientific community.

Overall, the coverage of references, citing publications, and author's publications have a comprehensive effect on citation, where references account for the most important role. It is crucial for authors to extensively select high-quality publications both inside and outside of their area of bibliometric as references. Diversified sources of references have the potential to influence other areas and contribute to the development of bibliometric research in the era of big data.

**Limitations**

We conducted the study on the influence of area coverage on citations based on bibliometric papers published between 2012 and 2021. However, the results may not be applicable to papers from other fields as research paradigms differ across disciplines, particularly between natural and social sciences. Therefore, we suggest including papers from various disciplines as samples in future studies.

**Funding**

Funded by Natural Science Foundation of China, NSFC (L2224027).

**Declarations**

The authors declare that they have no conflicts of interest with respect to the research, authorship and publication of this article and all the data is collected from open sources and analyzed by scientific software.